\documentclass[%
aip,amsmath,amssymb, reprint
]{revtex4-1}

\usepackage{graphicx}
\usepackage{dcolumn}
\usepackage{bm}
\usepackage[ruled,vlined]{algorithm2e}
\usepackage[utf8]{inputenc}
\usepackage[T1]{fontenc}
\usepackage{mathptmx}
\usepackage{listings}
\usepackage{amsmath}
\usepackage{leftidx}
\usepackage{tikz}
\usepackage{natbib}
\setcitestyle{square, comma, numbers,sort&compress, super}
\usepackage{cancel}
\usetikzlibrary{trees}
\usepackage{midfloat}
\usetikzlibrary{decorations.pathmorphing}
\usetikzlibrary{decorations.markings}
\usetikzlibrary{automata,positioning}
\usepackage{float}
\usepackage{caption}
\usepackage{simplewick}
\usepackage{csquotes}
\usepackage{graphicx}
\usepackage{subcaption}
\usepackage{multirow}
\usepackage{booktabs}
\usepackage{cleveref}
\usepackage{amsmath}
\usepackage{tabularx}
\usepackage{natbib}
\usepackage{rotating}
\usepackage{mathtools}
\usepackage{xspace}
\usepackage{colortbl}
\usepackage{xcolor}
\usepackage{todonotes}
\definecolor{codegreen}{rgb}{0,0.6,0}
\definecolor{codegray}{rgb}{0.5,0.5,0.5}
\definecolor{codepurple}{rgb}{0.58,0,0.82}
\definecolor{backcolour}{rgb}{0.95,0.95,0.92}
\lstdefinestyle{mystyle}{
    backgroundcolor=\color{backcolour},   
    commentstyle=\color{codegreen},
    keywordstyle=\color{magenta},
    numberstyle=\tiny\color{codegray},
    stringstyle=\color{codepurple},
    basicstyle=\ttfamily\footnotesize,
    breakatwhitespace=false,         
    breaklines=true,                 
    captionpos=b,                    
    keepspaces=true,                 
    numbers=left,                    
    numbersep=5pt,                  
    showspaces=false,                
    showstringspaces=false,
    showtabs=false,                  
    tabsize=2
}

\DeclareUnicodeCharacter{2212}{xxx Helloooo}

\begin{document}

\preprint{AIP/123-QED}

\author{Anish Chakraborty} \affiliation{Department of Chemistry, Indian Institute of 
Technology Bombay, Powai, Mumbai 400076, India} 
\author{Pradipta Kumar Samanta} \affiliation{Max Planck Institute for Solid State 
Research, D-70569 Stuttgart, Germany}
\author{Rahul Maitra}\email{rmaitra@chem.iitb.ac.in} 
 \affiliation{Department of Chemistry, Indian Institute of Technology Bombay, Powai, 
 Mumbai 400076, India}

\title{Accurate determination of excitation energy: An equation-of-motion
approach over a bi-exponential Coupled Cluster theory}

\begin{abstract}

The calculation of molecular excited states is critically important to decipher 
a plethora of molecular properties. In this manuscript, we develop an equation of
motion formalism on top of a bi-exponentially parametrized ground state wavefunction
towards the determination of excited states. While the ground state bi-exponential 
parametrization ensures an accurate description of the wavefunction through the
inclusion of high-rank correlation effects, the excited state is parametrized by a 
novel linear response operator with an effective excitation rank beyond two. 
To treat the ground and the excited states in the same footings, in addition 
to the conventional one and two-body response operators, we introduced 
certain two-body "generalized" response operators with an effective excitation rank 
of one. We introduce a projective formulation towards the determination of the 
perturbed amplitudes for the set of "generalized" operators. Our formulation
entails a significantly small number of unknown parameters and is shown to be highly 
accurate compared to allied methods for several difficult chemical systems.

\end{abstract}

\maketitle
\section{Introduction}

Molecular properties play a pivotal role in understanding
the nature of the chemical systems. These properties are commonly
obtained by observing the response of chemical systems by
an external field as a perturbation. Over the last few 
decades, several quantum chemical-based methods have been 
formulated to calculate the molecular energetics 
as well as the molecular properties in an effective manner 
to understand complicated chemical systems. Since it's 
inception, Coupled Cluster (CC)\cite{cc1,cc2,cc3,cc4} being 
a powerful tool for electronic structure correlation calculations, 
it is worthwhile to explore the efficacy of CC-based methods in 
calculating the molecular properties. 

Calculation of the excited states is not always straightforward 
and one may often face difficulties that originate
from the inadequate inclusion of ground-state correlation effects.
Insufficient inclusion of these effects renders a deficient 
description of the excited states and hence, a balanced
inclusion of the ground state correlation is important. Within
the single reference (SR) framework with SD excitations, the desired 
precision of the ground state wavefunction is often not achievable. 
One also needs to account for the orbital relaxation effects which 
are not properly accounted for in many theories.  
In general, for the calculations of first-order molecular property 
i.e. excitation energy, one may resort to two approaches as follows:
\begin{enumerate}
    \item The first approach involves the calculation of absolute 
    energies of all relevant excited states and subsequently determining 
    the excitation energies as the energy differences between 
    these states and the correlated ground state.
    \item The second approach entails the direct determination
    of excitation energy without explicitly determining the 
    relevant excited state energies.
\end{enumerate}
However, it is our contention that theories falling within the former 
category suffer from the insufficient treatment of orbital relaxation 
effects. Methods like multi-reference configuration interaction 
(MRCI)\cite{mrci} or complete active space second-order perturbation 
theory (CASPT2)\cite{mcscf} represent prominent examples of the first 
category. While these methods are capable of handling the multi-reference 
static correlation in an accurate manner, often exhibit limitations in 
terms of accuracy in calculating dynamical correlation. In the context 
of this manuscript, our primary focus will be on the methods that fall 
in the second category. These methods aim to directly determine the 
excitation energy spectrum, which has been an active area of 
research over the past couple of decades. This particular approach treats 
the ground and excited states on equal footing stemming the possibility 
of expressing the concerned excited states in terms of the ground state. 
This makes the calculation of energy difference feasible by canceling out 
the common correlation terms. The working principle of this scheme can 
be realized in a couple of ways i.e. either through time-dependent 
Linear Response (LR) formulation where excitation energy is calculated 
from pole structure or through factorized ansatz, where the correlation
effects are taken care of by the excited waveoperators. At the juncture 
of these two schemes, Malrieu \textit{et al.} developed a configuration 
interaction (CI) based methodology known as difference dedicated CI 
(DD-CI)\cite{ddci1,ddci2} which entails the construction of an effective 
Hamiltonian to determine the excitation energy in an efficient manner.
One may note that, in the context of SRCC theory, there are several 
other methods that encapsulate the idea pertaining to the second scheme. 
Mukherjee \textit{et al.}\cite{LR5} invoked a time-dependent formulation
to construct CC based linear response methodology to calculate spectroscopic
properties. Along the same line, Monkhrost\cite{LR1} and Dalgaard\cite{LR2}
introduced a time-dependent CC approach to calculate dynamic response properties.
Later on, Christiansen \textit{et al.}\cite{LR4} applied Fourier component 
variational perturbation theory for calculating the response functions. Also, 
a similar development was carried out by Koch and Jorgensen.\cite{LR3} 
Nakatsuji's formulation of symmetry-adapted cluster expansion based configuration
interaction (SAC-CI)\cite{sacci1,sacci2} towards the calculation of excited 
states is one important landmark in this avenue. Along a similar philosophy, 
Bartlett \textit{et al.} introduced equation-of-motion CC 
(EOM-CC)\cite{eom1,eom2,eom3,eom4} towards the calculation of excited state 
properties that have been extremely successful and widely used over the years.
It has been extensively observed that EOM-CCSD captures singly dominated
excited states with a high degree of precision, while often facing difficulties 
when the excited state is dominated by double excitations or the ground state 
is prone to degeneracy. One may invoke higher-rank dynamical correlation
effects in both ground and excited states to circumvent the problem. 
The EOM formulation with CCSDT\cite{EOM-CCSDT1,EOM-CCSDT2} and 
CCSDt\cite{eomccsdt1,eomccsdt2} takes care of the high-rank correlation effects 
in both ground and excited states, although they suffer from an elevated 
computational cost of iterative ($O(n^8)$) with a large number of unknown 
parameters $(O(n_o^3n_v^3))$. However, one may make judicious approximations 
towards an economical inclusion of high-rank excitations as done in several 
methods like EOM-CCSDT-n\cite{EOM-CCSDT-n1,EOM-CCSDT-n2}, 
EOM-CCSD(T)\cite{EOM-CCSDT-n1}, EOM-CCSD($\tilde{T}$)
\cite{EOM-CCSDT-n2}, EOM-CCSD(T’)\cite{EOM-CCSDT-n2} and CCSDR(3)
\cite{EOM-CCSDR}. Piecuch and co-workers introduced left-eigenstate completely 
renormalized equation-of-motion coupled cluster singles, doubles, 
and non-iterative triples (CR-EOMCCSD(2,3)).\cite{creomccsd} and its variants,
known as $\delta$-CR-EOMCC(2,3).\cite{delcreomcc1,delcreomcc2}
through the calculation of the tri-excited moment of EOM-CCSD. Both of these
methods have been extremely successful in studying the excited states
which are dominated by double excitations. CC3\cite{ex_cc3_1,ex_cc3_2,ex_cc3_3} 
methods have also been widely used as an accurate tool to calculate excitation energy
as they incorporate the effects of connected triples.
Krylov\cite{spinflip1,spinflip2} developed a spin-flip formulation 
of the EOM to extend the applicability towards the treatment of open-shell
systems starting from a single reference state. 
There are several methods that employ similarity transformation to 
determine the quantities of spectroscopic interests like IP, EA, and EE.
These methods rely on the solution of the ground state CC theory 
where the h-p and 2h-2p blocks of the effective Hamiltonian $e^{-T}He^T$
are iterated to zero. With the converged solution of the ground state 
wavefunction, one introduces a normal ordered exponential $\{e^{S}\}$ where
$S$ is the two-body operator with an effective hole-particle (hhhp or hppp) rank  
of one. The Fock Space CC theory 
(FSCC)\cite{fscc1,fscc2,fscc3,fscc4,fscc5,fscc6,fscc7,fscc8,fscc9,fscc10,fscc11,fscc12,fscc13,fscc14,fscc15,fscc16,fscc17,fscc18} 
introduces a second similarity transformation $G=\{e^{s}\}^{-1}He^S$ where 
each sub-block of $G$ corresponds to a definite excitation level and can be 
diagonalized separately. This amounts to the fact that the single excitation 
can be decoupled from double excitation. Following the notion of FSCC formalism, 
Nooijen proposed another methodology based on the similarity transformation 
of Hamiltonian for the determination of the excited state, known as Similarity 
Transformed EOM (STEOM)\cite{steom,steom1,steom2}. The STEOM approach 
introduces a set of active orbitals where the entire diagonalization procedure 
is cast as an eigenvalue problem within the active sub-block. Note that, apart 
from the contribution of the connected triple excitations, STEOM also possesses 
an implicit contribution from the \enquote{disconnected} triples, and this is 
later verified by Meissner and Bartlett via the dressing of the EOM-CC matrix to 
incorporate the effects of triple excitations.\cite{triple_ex_fscc1,triple_ex_fscc2} 
This compact diagonalization over a subspace enables STEOM to serve the dual purpose 
of high efficiency and low computational overhead. It is worthwhile to mention here 
that Mukherjee \textit{et al.} had shown that there is rigorous consanguinity 
between FSCC and IP/EA-EOM with STEOM\cite{fscc_steom_eom1,fscc_steom_eom2}.
A very interesting observation is that the solution of the ground state 
$t$-amplitudes and the $s$-amplitudes for the excited states are decoupled. 
An alternate approach can be introduced to solve a ground state with the inclusion
of high-rank correlation through a coupled solution of $s-$ and $t-$ 
amplitudes. Such a strategy has been developed by the present
authors via a bi-exponential parameterized wavefunction ansatz and it has been
extremely successful in solving the ground state energetics of weakly to
moderately strong correlated systems. The resulting method is
termed as iterative n-body excitation inclusive CC \textit{(iCCSDn)}
\cite{iccsdn1,maitra_coupled_2017,maitra_correlation_2017}.
One may ask a pertinent question: if the ground state is
solved through the coupled set of equations between $s-$ and $t-$ amplitudes,
whether one can determine the excited state with an equivalent parametrization.
In this work, towards the determination of the excited
states, we employ the same set of operators, akin to the ground state, within
the EOM-CC framework where the linear response operator includes the effective
triple excitations through a connected action over a certain \textit{contractible 
set of orbitals (CSO)} between two low-rank operators. The expansion of the rank-three 
response operator thus constructed is chosen to be of overall first
order. The contracted composite hole-particle structure of the response operator 
provide an ideal platform to formulate the theory with what may look like 
an EOM-CCSDT; however, instead of computing
the perturbed $T_3$ amplitudes which are $(O(n_o^3n_v^3))$, we invoke a perturbative 
expansion coupled with suitable decoupling conditions that lead us to 
bypass the redundancy towards
the determination of perturbed $s$ and $t$ amplitudes 
which enter into the response vector. A similar approach was recently 
introduced by the present authors\cite{iccsdls} where such parameter redundancy was
bypassed with a bi-variational approach; however, we now extend our
methodology towards the treatment of excited states where the coupling 
between the perturbed and the ground state vectors make it theoretically
challenging. We will refer to our methodology as EOM-iCCSDn.

In the subsequent sections, we first go through some of 
the key attributes of the ground-state iCCSDn methodology. This will be
followed by the formal development of EOM-iCCSDn where we will first cast the 
set of equations involving non-commuting operators in terms of 
commuting composite operators. We will then delineate our theoretical 
foundation to determine the perturbed amplitudes via the construction of a 
set of projection equations, followed by a physically motivated sufficiency 
condition. Towards the end of this section, we will also present the 
structure of the working equations and explain the terms that enter into it. 
Finally, we will present a series of numerical applications to
benchmark the efficacy of our proposed methodology and will compare
the results against some of the robust and state-of-the-art methods.

\section{Theoretical Development}
\subsection{A brief theoretical account of the ground state formulation:}

In the iCCSDn theory, the effect of high-rank dynamical 
correlation is included via low-rank waveoperator ansatz and towards
this, one introduces a set of \textit{scattering operators}, $S$ along with the 
standard cluster operators $T$. The scattering operators
have two components: an excitation component from 
occupied (hole) orbitals to the unoccupied (virtual)
orbitals and the other vertex contains scattering 
in either occupied or virtual space. The scattering 
vertex contains a single (quasi-) hole or (quasi-)
particle type one-electron state destruction operator 
and that divides $S$ into two parts i.e. $S_h$ (with 
quasi-hole destruction operator) and $S_p$ (with quasi-particle 
destruction operator), respectively. For instance, for a 
$S_h$ operator labeled by the Hartree-Fock (HF) occupied orbitals
$i,j$ and $m$ and virtual orbital $a$ with the associated normal 
ordered (with respect to HF vacuum) operator string 
$\{a^\dagger m^\dagger ji\}$ acts non-trivially 
on those selected set of determinants in which $i$ and $j$ are 
occupied while $a$ and $m$ are unoccupied. Similar analysis 
maybe done for the $S_p$ operators as well. In that sense, each 
component of the $S=S_h + S_p$ operators and their algebraic
expressions are given as:
\begin{eqnarray}
S_{h} = \frac{1}{2} \sum_{amij} s^{am}_{ij} \{a^\dagger m^\dagger ji\}
\end{eqnarray}
and
\begin{eqnarray}
S_{p} = \frac{1}{2} \sum_{abie} s^{ab}_{ie} \{a^\dagger b^\dagger ei\} 
\label{eqxx1}
\end{eqnarray}
Here, \textit{a, b, c, ...,e,...} etc. denote the set
of particle orbitals and \textit{i, j, k, ...,m,...} etc.
are the set of hole orbitals with respect to the HF
vacuum. Note that, in the definition of $S_{h}$ and $S_{p}$, 
\enquote*{\textit{m}} and \enquote*{\textit{e}}
respectively, appear as quasi-hole and quasi-particle
destruction operator and together, they will be said to
form the CSO. We will demonstrate later that the accuracy of our results are
somewhat sensitive to the dimensions of CSO.
Note that due to the built-in projector in the definition
of $S$, its action on HF reference is trivially zero. 
This may be referred to as the vacuum annihilating 
condition (VAC): $S_h|\Phi\rangle = 
0;\hspace{0.3cm} S_p|\Phi\rangle = 0$. It is important
to note here that the effect of higher-order excitations
is subsumed by the selective action of $S$ operators
on doubly excited functions, which is realized 
via the contraction between $S$ and $T_2$ operators.

The destruction operator in $S$ restricts them 
from commuting with the cluster operators, $T$, with which it shares 
at least one orbital index of CSO. Also, the various components of $S$ do not commute among
themselves. Additionally, the projector ensures that the operator acts non-trivially 
on selected excited determinants that are generated by the preceding 
action of a $T$ operator. These cluster operators 
contain at least one orbital index of CSO and hence they
are non-commutative with those $S$ and this non-commutativity
is the key to the genesis of the triply excited determinants
which is conceptualized to be generated by the action 
of the $S$ operator on the doubly excited functions.

With the introduction of the scattering operators, one may write the waveoperator 
as a product of two separate exponential operators as:
\begin{equation}
    \Omega = \{e^{s}\} e^{T}
\label{eqxx2}
\end{equation}
For the details of the genesis of this particular 
waveoperator form, we refer to our earlier 
works\cite{iccsdn1,maitra_correlation_2017}
in this context. 

In Eq. \ref{eqxx2}, $\{...\}$ denotes the normal 
ordering with respect to HF reference which prevents
the $S-S$ contraction, resulting in a naturally 
truncating effective Hamiltonian (\textit{vide infra}).
Note that the cluster operator $T$ consists of 
one and two-body operators ($T=T_1+T_2$) and thus in the
iCCSDn theory, one mimics the connected higher body 
correlation effects by restricting oneself to one and 
two-body parametrization.

On an operational level, for the many-body implementation
of the double exponential iCCSDn theory,
the solutions for the optimized $s-$ and $t-$amplitudes (corresponding to 
$S$ and $T$ operators, respectively) are determined in a coupled manner by
working entirely in the operator space. Note that, due to the
non-triviality of the similarity transformation of normal ordered
operator, one can define the first similarity transformed effective Hamiltonian, 
$W$ such that
\begin{equation}
\{e^S\} W = H \{e^S\}
\label{eqxx4}
\end{equation}
is satisfied. Since the inverse operation of a normal ordered ansatz is not
explicitly defined, one may determine $W$ with a recursive substitution 
technique\cite{iccsdn1}. The resulting structure of $W$ then 
takes the form:

\begin{eqnarray}
W=\{\contraction{}{H}{}{e^S} H e^S\}  - \{\contraction[2ex]
{}{(e^S-1)}{}{\contraction{}{H}{}{e^S} H e^S} (e^S-1)
\contraction{}{H}{}{e^S} H e^S\} +
\{\contraction[3ex]{}{(e^S-1)}{}{\contraction[2ex]
{}{(e^S-1)}{}{\contraction{}{H}{}{e^S} H e^S} (e^S-1)
\contraction{}{H}{}{e^S} H e^S} (e^S-1) \contraction[2ex]
{}{(e^S-1)}{}{\contraction{}{H}{}{e^S} H e^S} (e^S-1)
\contraction{}{H}{}{e^S} H e^S \}  - \cdot\cdot\cdot 
\label{eqxx6}
\end{eqnarray}
Where $W$ is an effective many-body operator and various truncation
schemes upon this will give rise to certain approximations. Note that,
in our scheme, we will truncate $W$ after the second term on the
right hand side of Eq.\ref{eqxx6}. It is important to emphasize
here that, irrespective of the truncation scheme, there is at least one 
destruction operator arising out of the contraction between the
Hamiltonian and S from the right. Now, after this judicious truncation
upon $W$, one can carry out the second similarity transformation
on top of it resulting in the construction of effective
Hamiltonian, $R=e^{-T}We^{T}$. Through the many-body expansion of
this double similarity transformed Hamiltonian, $R$ can be written as:

\begin{equation}
R  = r_0 + r_p^q\{E_q^p\} + \frac{1}{4} r_{pq}^{st}\{E^{pq}_{st}\} + 
\frac{1}{36} r_{pqs}^{tuv}\{E^{pqs}_{tuv}\} + ...
\label{eqxx7}
\end{equation}
where $p,q,r,s,...$ are general (hole or particle)
orbital indices. Clearly, the above many-body 
expansion contains all $N$-body terms with all 
possible hole-particle scattering structures. 
Following Nooijen\cite{nooijen2000}, the 
corresponding $t-$ and $s-$ amplitudes are 
obtained by demanding the associated amplitudes
vanish:
\begin{equation}
r^{a}_{i} = r^{ab}_{ij} = r^{am}_{ij} = r^{ab}_{ie} = 0
\label{eqxx18}
\end{equation}
Where the first two terms are corresponding to the determination of one and 
two-body cluster amplitudes whereas the later two terms are for
determination of amplitudes corresponding to $S_h$ and $S_p$ respectively. 
It is important to note here that, the construction of this effective Hamiltonian, 
$R$ and optimization of $t-$ and $s-$ amplitudes are done in an iterative
manner. The high-rank correlation effects are incorporated via
the contraction of a few contractible orbitals between $S$ and $T_2$
instead of using explicit high rank $T_3$ operator resulting in the
decrease of the overall computational scaling comparable to CCSD.
In the subsequent section, we will demonstrate the formal development
of our EOM-iCCSDn methodology.

\subsection{EOM treatment of iCCSDn towards excitation energy:}
In this section, we will proceed towards the conceptualization 
of a linear response operator in terms
of the one and two-body excitation operators as well as the 
scattering operators. While the determination of the
perturbed one and two-body operators can be done in a 
standard projective formulation, due to the vacuum 
annihilating condition, no such closed projective 
form of matrix equation can be derived for the scattering
operators, $S$. In order to bypass the complications
arising due to the VAC of $S$, we will form a composite
through contraction of $S$ and $T$ such that the resulting effective operator 
having standard hole-particle excitation structure. This will help us do to 
mathematical manipulations towards the development of 
the EOM-iCCSDn theory in terms of a set of commuting 
effective operators.

\subsubsection{Casting the problem in terms of commuting operators: 
the genesis of the EOM-iCCSDn ansatz}

Towards the development of the EOM-iCCSDn ansatz, we first
demonstrate that the ground state double exponential 
ansatz can be expressed in terms of hole-particle
excitation like connected effective operators. This
would also help us in the conceptualization of a linear
response operator that can have a simple commutation 
property with the ground state excitation operators.

Noting the fact that the waveoperator consists of a 
product of two exponential operators in normal
ordering, one may explicitly apply Wick's theorem to
write Eq. \ref{eqxx2} as
\begin{equation}
    \Omega = \{e^S e^{\contraction{}{S}{}{T} S T} e^T\} 
    \label{eqmm8}
\end{equation}
Here the term ${\contraction{}{S}{}{T}} ST$ denotes 
all the possible single, double, triple, $...$ 
contractions between various powers of $S$ and $T$. Note that the 
quantities $e^S$ and $e^T$ inside the normal ordering are the parts 
coming out of the operator which is not explicitly connected. 
Also, due to the normal ordering, there is 
no explicit contraction between two $S$ operators. One may 
note that all these $S$ operators which are not explicitly 
connected also forms an exponential series, owing to 
the property of an exponential operator.

The quantity ${\contraction{}{S}{}{T} S T}$, 
by construction, has the hole-particle structure 
like an excitation operator, the lowest of 
which starts from rank three. Due to the similar 
hole-particle excitation structure between 
$T$ and ${\contraction{}{S}{}{T} S T}$, they commute. 
Thus Eq. \ref{eqmm8} can equivalently be written as:
\begin{equation}
     \Omega = \{e^S e^{\contraction{}{S}{}{T} S T + T}\}
    \label{eqmm9}
\end{equation}
The $e^S$ operator which is not explicitly connected has the 
free (uncontracted) destruction operator index that 
destroys the HF vacuum when the waveopeartor acts upon it. 
Thus, the action of the waveoperator $\Omega$ can be written 
in a simplified manner as:
\begin{equation}
    \Omega |\phi_{0}\rangle = \{e^S e^{\contraction{}{S}{}{T} S T + T}\} 
    |\phi_{0}\rangle = \{ e^{\contraction{}{S}{}{T} S T + T}\} |\phi_{0}\rangle
\label{eqmm10}
\end{equation}
Since $(T + \contraction{}{S}{}{T} S T)$ forms a set of 
mutually commuting operators that have hole-particle
excitation structure, one may remove the normal ordering
such that Eq.\ref{eqmm10} may equivalently be written as:
\begin{equation}
    |\psi_{0} \rangle = e^{\contraction{}{S}{}{T} S T + T}|\phi_{0}\rangle
    \label{eqmm300}
\end{equation}
For notational simplicity, we would subsequently write 
$\contraction{}{S}{}{T} S T + T$ as $Z$ which has the 
various ranks of hole-particle excitation structure.

In order to work with a commuting set of operators, we 
would now design the linear response operator for the $k^{th}$ 
excited state, $R^k$ in a similar manner such that it commutes 
with $Z$. Such an operator can be defined as:
\begin{equation}
R^{k} = R_{1}^{k} + R_{2}^{k} + 
{\contraction{}{R_{S}^{k}}{}{Y_{2}} R_{S}^{k} Y_{2}}
    \label{eqmm11}
\end{equation}
Here $R_{1}^{k}$ and $R_{2}^{k}$ have the one and two body 
hole-particle excitation structure. The term 
${\contraction{}{R_{S}^{k}}{}{Y_{2}} R_{S}^{k} Y_{2}}$ denotes a
three-body linear response operator in terms of the first-order 
variation of $S$ operators. All the operators employed in Eq.\ref{eqmm11} can be 
written explicitly as: $R_{1}^{k} = r_{1}^{k}\Hat{Y_{1}}$;\hspace{0.05cm} 
$R_{2}^{k} =r_{2}^{k}\Hat{Y_{2}}$;\hspace{0.05cm} $R_{S}^{k} 
= r_{s}^{k} \Hat{Y_{S}}$, where $\Hat{Y}_{..}$ denotes 
the associated string of creation and annihilation operators, 
and $r_{...}$ denotes the associated perturbed amplitudes corresponding 
to the operators. Note again that ${\contraction{}{R_{S}^{k}}{}{Y_{2}} R_{S}^{k} Y_{2}}$ 
is a three-body hole particle excitation operator and hence $R^k$ commutes with 
$Z$. This would help us write $k^{th}$ excited state in terms of a set of commuting
operators such that all the mathematical manipulations can be done in a simplified 
manner like that of a conventional EOMCC method. 
\begin{equation}
    |\psi_k\rangle = R^{k}e^{Z}|\phi_{0}\rangle
    \label{eqmm12}
\end{equation}
We reiterate here that due to working with the commuting operators 
$R^k$ and $Z$ with hole-particle excitation structure with respect to 
the HF reference, we have removed the explicit use of the 
normal ordering in $e^Z$.

\subsubsection{Determination of perturbed amplitudes: projection formalism for excited 
state}

We now proceed towards the development of EOM-iCCSDn 
formulation for the many-body excited states. We will 
derive the working equations for determining the 
perturbed $T$ and $S$ amplitudes. Note that the ground
state is now described in terms of $Z=\contraction{}{S}{}{T} S T + T$
that has an excitation structure.
Following the conventional methodology, we start with the 
Schr{\"{o}}dinger equation for the $k^{th}$ excited state, $\psi_k$:
\begin{equation}
    HR^{k}e^{Z}|\phi_{0}\rangle = E_{k}R^{k}e^{Z}|\phi_{0}\rangle
    \label{eqmm13}
\end{equation}
Due to the inherent commutativity property between $R^k$ and 
$Z$, one may rewrite Eq.\ref{eqmm13} in the following manner:
\begin{equation}
    He^{Z}R^{k}|\phi_{0}\rangle = E_{k}e^{Z}R^{k}|\phi_{0}\rangle
    \label{eqmm130}
\end{equation}
Upon multiplying both sides of Eq.\ref{eqmm130} by $e^{-Z}$ we can get:
\begin{eqnarray}
    e^{-Z}He^{Z}R^{k}|\phi_{0}\rangle = E_{k}e^{-Z}e^{Z}R^{k}|\phi_{0}\rangle \nonumber \\
    \implies H_{eff}R^{k}|\phi_{0}\rangle = E_{k}R^{k}|\phi_{0}\rangle
    \label{eqmm131}
\end{eqnarray}
Here, $H_{eff}=e^{-Z}He^{Z}$ denotes the zeroth order effective Hamiltonian.
Simultaneously, multiplication of $R^{k}$ on the Schr{\"{o}}dinger equation '
for the ground state yields:
\begin{eqnarray}
    R^{k}He^{Z}|\phi_{0}\rangle = E_{0}R^{k}e^{Z}|\phi_{0}\rangle
    \label{eqmm132}
\end{eqnarray}
Subsequently, a few algebraic alterations upon Eq.\ref{eqmm132} generate 
the following discerning equation:
\begin{eqnarray}
    R^{k}H_{eff}|\phi_{0}\rangle = E_{0}R^{k}|\phi_{0}\rangle
    \label{eqmm133}
\end{eqnarray}

\noindent Subtracting Eq.\ref{eqmm133} from Eq.\ref{eqmm131},
one may now write the EOM equation in terms of the matrix eigenvalue equation as follows:
\begin{equation}
    [H_{eff},R^{k}]|\phi_{0}\rangle = \Delta E_{k}R^{k}|\phi_{0}\rangle
    \label{eqmm14}
\end{equation} 
Here, $\Delta E_{k} = (E_{k} - E_{0})
$, which denotes the excitation energies
corresponding to $k^{th}$ excited state. In order to determine the perturbed 
one body amplitudes ($r_{1}^{k}$), we project Eq. \ref{eqmm14} by the singly 
excited determinants, $\chi_s$ as follows:
\begin{eqnarray}
\langle \chi_s | [H_{eff},R^{k}] | \phi_{0}\rangle = \Delta E^{k}
\langle \chi_s | R^{k} | \phi_{0}\rangle \nonumber \\ \implies
\langle \phi_{0} | Y_{1}^{\dagger} [H_{eff},R^{k}] | \phi_{0}
\rangle = \Delta E_{k} r_{1}
\label{eqmm15}
\end{eqnarray}
Here $\chi_s$ is the singly excited determinant which 
can be expressed as $\langle \chi_s| = \langle \phi_{0}|Y_{1}^{\dagger}$.
Note that, for the brevity of the notation, we will denote
$r_{x}^{k}$ as $r_{x}$ from now onwards where $x=1,2,s$.
Expanding $R^k$, one may further rewrite Eq.\ref{eqmm15} in
a long hand notation as:
\begin{eqnarray}
\langle \phi_{0} | Y_{1}^{\dagger} [H_{eff},Y_{1}] | \phi_{0}
\rangle r_{1} + \langle \phi_{0} | Y_{1}^{\dagger} [H_{eff},Y_{2}] | 
\phi_{0} \rangle r_{2} + \nonumber \\
\langle \phi_{0} | Y_{1}^{\dagger} 
[H_{eff},\contraction{}{Y_{s}}{}
{Y_{2}} Y_{S} Y_{2}] | \phi_{0} \rangle r_{S} = \Delta E_{k} r_{1}
\label{eqmm16}
\end{eqnarray}
In a similar manner, the equation to determine $r_{2}$ 
can be explicitly written by projecting Eq.\ref{eqmm14}
by doubly excited determinants, $\langle \chi_d | = 
\langle \phi_{0}| Y_{2}^{\dagger}$ in the following manner:
\begin{eqnarray}
\langle \phi_{0} | Y_{2}^{\dagger} [H_{eff},Y_{1}] | \phi_{0}
\rangle r_{1} + \langle \phi_{0} | Y_{2}^{\dagger} [H_{eff},Y_{2}] | 
\phi_{0} \rangle r_{2} + \nonumber \\
\langle \phi_{0} | Y_{2}^{\dagger} 
[H_{eff},\contraction{}{Y_{s}}{}
{Y_{2}} Y_{S} Y_{2}] | \phi_{0} \rangle r_{S} = \Delta E_{k} r_{2}
\label{eqmm17}    
\end{eqnarray}
The $Y_{1}^{\dagger}$ and $Y_{2}^{\dagger}$ are the 
one-body and two-body de-excitation operators. At this
stage, we point to the non-trivial aspect of our EOM-iCCSDn
formulation. While the one and two-body perturbed amplitudes can be
computed in a straightforward manner, there is no such a closed form 
of equation possible to determine $r_{s}$ 
due to the killer conditions (or VAC). This implies that there is no
such bra projection by which one may get the $r_{s}$ 
amplitudes directly. However, an equation for 
$r_{s}$-amplitudes can be constructed from three-body 
excitation blocks as we discuss below. One may project
Eq. \ref{eqmm14} by the triply excited determinants 
$\langle \chi_{t_I} |= \langle \phi_{0} | Y_{3_I}^\dagger$ 
to arrive at:
\begin{eqnarray}
\langle \chi_{t_I} |  [H_{eff},R^{k}]_{I} | \phi_{0}\rangle = \Delta E_{k}
\langle \chi_{t_I} | R^{k}_{I} | \phi_{0}\rangle \nonumber \\
\langle \phi_{0} | Y_{3_I}^{\dagger} [H_{eff},R^{k}]_{I} | \phi_{0} \rangle =
\Delta E_{k} \langle \phi_{0} | Y_{3_I}^{\dagger} R^{k}_{I} | \phi_{0}
\rangle
\label{eqmm18}
\end{eqnarray}
where $I$ denotes the collective three hole - three particle orbital labels. 
In a long-hand notation, Eq. \ref{eqmm18} can be 
written as:
\begin{eqnarray}
\langle\phi_{0} | {Y_{3_I}^\dagger} [H_{eff},R_{1}]_{I} | \phi_{0}\rangle + 
\langle\phi_{0} | {Y_{3_I}^\dagger} [H_{eff},R_{2}]_{I} | \phi_{0}\rangle \nonumber \\
+ \langle\phi_{0} | {Y_{3_I}^\dagger} [H_{eff}^{J}, \sum_{\alpha,M}
(\contraction{}{R_{s_\alpha}}{}{Y_{2_M}} R_{S_\alpha} Y_{2_M})_{L}]_{I} 
|\phi_{0}\rangle \nonumber \\
= \Delta E_{k} \langle\phi_{0} | {Y_{3_I}^\dagger} \sum_{\beta,M^{\prime}}
(\contraction{}{R_{s_\beta}}{}{Y_{2_{M^{\prime}}}} R_{s_\beta} Y_{2_{M^{\prime}}})_{I} 
|\phi_{0}\rangle 
\label{eqmm21}
\end{eqnarray}
Here $\alpha$ is the combined orbital indices associated with operator $S$.
The indices $I,J,L,M$ are the combined indices for excitation-types orbital 
structures associated with various operators.
In the left hand side, the summation over the indices $\alpha$ and $M$ is 
restricted to certain orbital index tuples which upon contraction gives rise to 
3 hole - 3 particle excitation index tuple $L$, which subsequently gets 
contracted with $H_{eff}^J$ to produce an excitation level $I$.
It is worthwhile to mention here
that while we formulated this equation by explicitly projecting against the 
triply excited determinants, our unknown parameters are amplitudes for perturbed
$S$ operators. Although the perturbed amplitudes for $T_2$ operators can be
determined via Eq.\ref{eqmm17}, establishing a projective form of the equation for 
determining the perturbed $s$-amplitudes directly is not feasible. This limitation 
stems from the inability to compute the matrix elements (relative to $\phi_{HF}$) 
of an effective Hamiltonian resembling the structure of $S$ due to the VAC.
Moreover, the number of such three-body projections is far too many than
the number of $S$ and $T$ combined. This implies that several sets of $Y_2$
and $Y_S$ give rise to the same three-body excitation 
function. In other words, the three-body excitation manifold is redundant. To
remove such a redundancy, we resort to a \textit{physically motivated 
sufficiency condition} as explained below.

As previously mentioned, there are several pairs of indices $J$ 
and $L$ that give rise 
to an index tuple $I$. On the other hand, a specific 3h-3p index $L$ may be generated
by several combinations of $\alpha,M$. Let us pick a specific $\alpha$ that may 
get contracted with various $Y_{2_M}$ to generate the set of three-body excitation 
index tuple $L$. In other words, different $L$'s are generated through the variation 
of the index tuples solely in $\alpha$. Thus one may remove the summation over $\alpha$
in Eq. \ref{eqmm21} to arrive at:
\begin{eqnarray}
\langle\phi_{0} | {Y_{3_I}^\dagger} [H_{eff},Y_{1}]_{I} | \phi_{0}\rangle r_{1} + 
\langle\phi_{0} | {Y_{3_I}^\dagger} [H_{eff},Y_{2}]_{I} | \phi_{0}\rangle r_{2} \nonumber \\
+ \langle\phi_{0} | {Y_{3_I}^\dagger} [H_{eff}^{J}, \sum_{M}
(\contraction{}{Y_{s}}{}{Y_{2_\beta}} Y_{S} Y_{2_M})_{L}]_{I} |\phi_{0}\rangle 
r_{s} \nonumber \\ 
= \Delta E_{k} \langle\phi_{0} | {Y_{3_I}^\dagger} \sum_{M^{\prime}}
(\contraction{}{Y_{s}}{}{Y_{2_{M^{\prime}}}} Y_{s} Y_{2_{M^{\prime}}})_{I} |\phi_{0}\rangle r_{s} 
\nonumber \\
\hspace{0.5cm}
\forall (\alpha,\beta) \in \left\{r_{s_h},r_{s_p}\right\}
\label{eqmm22}
\end{eqnarray}
This sufficiency condition thus removes the redundancy by ensuring 
each unique $r_s$ amplitude is determined via a (weighted) sum 
of certain two-body operators. While The Eq.\ref{eqmm22} determines perturbed 
$s-$amplitudes, Eqs.\ref{eqmm16} and \ref{eqmm17} effectively 
determines the perturbed $t_1$ and $t_2-$ amplitudes. One may note here that
the sufficiency condition for determining the $r_s$ amplitudes allows one
to write the equations for determining the perturbed amplitudes 
($r_{1},r_{2}$ and $r_{s}$) in terms of a matrix eigenvalue equation 
as shown below:

\begin{equation}
\begin{pmatrix}
A_{Y_1 r_1} & B_{Y_1 r_2} & C_{Y_1 r_s} \\
D_{Y_2 r_1} & E_{Y_2 r_2} & F_{Y_2 r_s} \\
G_{Y_3 r_1} & H_{Y_3 r_2} & I_{Y_3 r_s}
\end{pmatrix}
\begin{pmatrix}
r_{1} \\
r_{2} \\
r_{s}
\end{pmatrix}
=
\Delta E_{k}
\begin{pmatrix}
r_{1} \\
r_{2} \\
r_{s}
\end{pmatrix}
\label{eqmm25}
\end{equation}
All the elements corresponding to the matrix are listed in table \ref{tabmm1}
in a detailed manner.
\begin {table}[htbp]
\centering
 \hglue -0.7cm
\setlength{\tabcolsep}{4.2pt}
\renewcommand{\arraystretch}{2.0}
\begin{tabular}{ c c }
\hline
\hline
Abbreviations & Matrix Representation  \\
\hline 
$A_{Y_1 r_1}$ & $\langle \phi_{0} | Y_{1}^{\dagger} [H_{eff},Y_{1}] | 
\phi_{0} \rangle$ \\
$B_{Y_1 r_2}$ & $\langle \phi_{0} | Y_{1}^{\dagger} [H_{eff},Y_{2}] | 
\phi_{0} \rangle$ \\
$C_{Y_1 r_s}$ & $\langle \phi_{0} | Y_{1}^{\dagger} [H_{eff},
\contraction{}{Y_{s}}{}{Y_{2}} Y_{S} Y_{2}] | \phi_{0} \rangle$ \\
$D_{Y_2 r_1}$ & $\langle \phi_{0} | Y_{2}^{\dagger} [H_{eff},Y_{1}] | 
\phi_{0} \rangle$ \\
$E_{Y_2 r_2}$ & $\langle \phi_{0} | Y_{2}^{\dagger} [H_{eff},Y_{2}] | 
\phi_{0} \rangle$ \\
$F_{Y_2 r_s}$ & $\langle \phi_{0} | Y_{2}^{\dagger} [H_{eff},
\contraction{}{Y_{s}}{}{Y_{2}} Y_{S} Y_{2}] | \phi_{0} \rangle$ \\
$G_{Y_3 r_1}$ & $\langle\phi_{0} | {Y_{3_I}^\dagger}
[H_{eff},Y_{1}]_{I} | \phi_{0}\rangle$ \\
$H_{Y_3 r_2}$ & $\langle\phi_{0} | {Y_{3_I}^\dagger}
[H_{eff},Y_{2}]_{I} | \phi_{0}\rangle$ \\
$I_{Y_3 r_s}$ & $\langle \phi_{0} | {Y_{3_I}^\dagger}
[H_{eff}^{J},\sum_{M}
(\contraction{}{Y_{s}}{}{Y_{2_M}} Y_{S} Y_{2_M})_{L}]_{I} |\phi_{0}\rangle$ \\
\hline
\hline
\end{tabular}
\caption{Detailed tensor expressions of all the matrix elements involved in the
 EOM-iCCSDn matrix eigenvalue equation}
\label{tabmm1}
\end{table}
One may note that the equations for $r_{s}$ are determined
through the three-body projection. However, the decoupling condition allows 
us to determine the $r_{s}$ and $r_{2}$ amplitudes separately, which
together are orders of magnitude less than that the typical number
of three-body equations, leading to enormous savings in the 
computational operation and memory. The matrix eigenvalue equation is 
implemented using our in-house code where we employed Davidson
algorithm\cite{davidson,davidson2} to solve for the lowest eigen roots
corresponding to each symmetry state.

\section{Results and discussion}
In this section, we will discuss about the accuracy exhibited 
by our methodology and compare the results against EOM-CCSD as well
as robust CC3 methodologies. Through several pilot numerical
applications on several challenging systems, we will demonstrate the precision
in predicting different single as well as double excitations dominated
states and the effect of the contractible set of orbitals. All the 
calculations were done using an in-house software platform which is 
interfaced to PySCF\cite{pyscf} that generates the integrals and orbitals.

\subsection{Carbon Monoxide (CO)}
Carbon Monoxide (CO) is one of the most widely used molecules for 
assessing the performance of any excited state methodology. Although 
$CO$ molecule belongs to $C_{\infty V}$ point group, but on an 
operational level, we will take the highest abelian group i.e. $C_{2v}$ and
Owing to that, there are four irreducible
representations of $A_{1},B_{1},B_{2}$ and $A_{2}$ which are used
to generate $\Sigma^{+},\Pi$,$\Sigma^{-}$ and $\Delta$ states 
in $C_{\infty V}$ point group. We have
calculated the excitation energy for $CO$ in three different basis
sets i.e. cc-pVDZ, cc-pCVDZ and TZVP and in all cases, all
the electrons were correlated. The results of all the calculations
are enlisted in table \ref{tabmm2}. For all the states studied in this 
work using cc-pVDZ basis set, the EE values predicted 
by EOM-CCSD and CC3 vary between 0.008 eV (for $1\Delta$) to 0.145 eV 
(for $3\Sigma^{+}$). For almost all these roots, the EOM-iCCSDn results 
fall in between the corresponding EOM-CCSD and CC3 values; with the lowest
EE difference (with respect to CC3) is of 0.004 eV for $1\Delta
(\pi-\pi^{*})$ and the highest difference is 0.119 eV for $3\Sigma^{+} 
(\sigma-\sigma^{*})$. 
\begin{sidewaystable*}[htbp]
\centering
\setlength{\tabcolsep}{2.65pt}
    \renewcommand{\arraystretch}{2.5}
    \vspace*{26em}
    \begin{tabular}{ccc>{\columncolor[gray]{0.8}}c>{\columncolor[gray]{0.8}}c
    >{\columncolor[gray]{0.8}}ccc>{\columncolor[gray]{0.8}}c>{\columncolor[gray]{0.8}}c>
    {\columncolor[gray]{0.8}}ccc>{\columncolor[gray]{0.8}}c>{\columncolor[gray]{0.8}}c
    >{\columncolor[gray]{0.8}}c>{\columncolor[gray]{0.8}}c}
    \hline
    \hline
    \multirow{4}{*}{States} & \multicolumn{16}{c}{Basis Sets} \\\cline{2-17}
                            & \multicolumn{5}{c}{\textbf{cc-pVDZ}} 
            & \multicolumn{5}{c}{\textbf{cc-pCVDZ}} & \multicolumn{6}{c}{\textbf{TZVP}} 
            \\
            \cmidrule(l){2-6} \cmidrule(l){7-11} \cmidrule(l){12-17}
                            & EOM-CCSD & CC3 & \multicolumn{3}{c}{EOM-iCCSDn}
            & EOM-CCSD & CC3 & \multicolumn{3}{c}{EOM-iCCSDn} & EOM-CCSD 
            & CC3 & \multicolumn{4}{c}{EOM-iCCSDn} \\
                            &     &    & (1,1) & (2,2) & (3,3)
            &     &          & (1,1) & (2,2) & (3,3) &  
            &          & (1,1) & (2,2) & (3,3) & (4,4) \\
    2$\Sigma^{+}$ & 15.922 & 15.806 & 15.925 & 15.908 & 15.880 & 15.916 & 15.792 & 15.920 & 15.904 & 15.876 & 13.214 & 13.176 & 13.188 & 13.188 & 13.202 & 13.206 \\
    3$\Sigma^{+}$  & 18.180 & 18.035 & 18.132 & 18.110 & 18.121 & 18.172 & 18.019 & 18.125 & 18.102 & 18.111 & 14.612 & 14.470 & 14.612 & 14.607 & 14.603 & 14.570 \\
    \hline
    1$\Pi$ & 10.275 & 10.199 & 10.262 & 10.250 & 10.235 & 10.292 & 10.214 & 10.279 & 
    10.267 & 10.252 & 10.100 & 10.001 & 10.092 & 10.081 & 10.064 & 10.052 \\
    2$\Pi$ & 15.885 & 15.803 & 15.819 & 15.813 & 15.813 & 15.901 & 15.812 & 15.835 & 
    15.829 & 15.829 & 14.971 & 14.818 & 14.953 & 14.947 & 14.941 & 14.929 \\
    \hline
    1$\Sigma^{-}$ & 13.259 & 13.251 & 13.255 & 13.266 & 13.279 & 13.260 & 13.250 & 13.256 & 13.267 & 13.280 & 13.110 & 13.104 & 13.105 & 13.115 & 13.131 & 13.139 \\
    \hline
    1$\Delta$ & 13.386 & 13.349 & 13.359 & 13.357 & 13.367 & 13.390 & 13.348 & 
    13.362 & 13.361 & 13.371 & 13.214 & 13.176 & 13.188 & 13.188 & 13.201 & 13.205 \\ 
    \hline
    \hline
    \end{tabular}
    \caption{Vertical excitation energies of $CO$ in various basis sets
    i.e. cc-pVDZ, cc-pCVDZ and TZVP. While most of the roots are single excitations
    dominated, $2B_{1}$ and $2B_{2}$ roots exhibit weak double excitation character.
    Note that, all the values given in the table are in eV. The values in the
    parenthesis denotes the number of occupied and virtual orbitals considered
    to include in CSO.}
    \label{tabmm2}
\end{sidewaystable*}
However, their actual magnitudes have a non-monotonic dependence on the choice 
of the CSO. Note that, except for $2\Pi (\sigma^{*}-\pi^{*})$ which has 
a weakly double excitation dominated character, all other states exhibit single 
excitation character. While the $3\Sigma^{+}$ root is purely dominated by single 
excitations as predicted by EOM-CCSD, interestingly, in EOM-iCCSDn, one 
finds a large component of $S$ ($S_{p}: 6b_{2}9b_{2}\rightarrow 
8b_{1}8b_{1}$, $7a_{1}10a_{1}\rightarrow8b_{1}8b_{1}$ and $S_{h}: 
7a_{1}7a_{1}\rightarrow9b_{2}6b_{2}$). This along with
a couple of large $T_2$ ($7a_{1}7a_{1}\rightarrow9b_{2}6b_{2}$) 
in turn, brings in a substantial effect of 
$T_3$ which otherwise cannot be captured through CC3. A similar trend 
can be observed for cc-pCVDZ basis where the difference of the 
EE values predicted by EOM-CCSD and CC3 vary between 0.01 eV 
(for $1\Delta$) to 0.153 eV (for $3\Sigma^{+}$). EOM-iCCSDn further 
lowers the EE error with the lowest error being 0.006 ev (for 
$1\Delta$) and the highest is 0.128 ev (for $3\Sigma^{+}$). Note that, 
similar to the cc-pVDZ basis, in cc-pCVDZ basis, $3\Sigma^{+}$ root 
shows higher contribution of $S$ operators which amounts to higher 
contribution from $T_3$.

The estimation of EE using a TZVP basis shows a somewhat different trend than 
the previous two cases. In this case, the lowest and highest EE difference 
for EOM-CCSD (with respect to CC3) is observed for $1\Delta$ (0.006 eV) 
and $2\Pi$ (0.153 eV), respectively. However, for EOM-iCCSDn, the highest 
error in EE is observed for $3\Sigma^{+}$ ($EE_{EOM-iCCSDn}-EE_{CC3}= 0.142$ eV).
In this case (as well as for $2\Sigma^{+}$ and $2\Pi$), several $S$ operators 
attain a significantly large magnitude.

\subsection{Formaldehyde}
Formaldehyde, being one of the simplest carbonyl compounds, undergoes 
lots of theoretical studies in the domain of excited state calculations.
The geometry of formaldehyde used in this calculation was taken from 
Thiel test set\cite{thiel}. The ground state electronic configuration 
for formaldehyde is:
\begin{equation}
    1a_{1}^{2}2a_{1}^{2}3a_{1}^{2}4a_{1}^{2}1b_{2}^{2}5a_{1}^{2}
    1b_{1}^{2}2b_{2}^{2}
\end{equation}
Similar to $CO$ molecule, owing to the $C_{2v}$ point group,
Formaldehyde has four irreducible representations namely 
$A_{1},B_{1},B_{2}$ and $A_{2}$. The excitation energies for
formaldehyde are calculated in different basis sets i.e. cc-pVDZ
and TZVP and all the results are enlisted in table \ref{tabmm3}.
Note that, for all the calculations, all the electrons were correlated.
For EOM-CCSD methodology, in the cc-pVDZ basis, the EE fluctuates between
0.018 eV (for $1A_{2}$) and 0.236 eV (for $2A_{1}$) with respect to CC3. 
For all the states, EE predicted by our EOM-iCCSDn methodology resides 
in between EOM-CCSD and CC3 values. The EE difference between the CC3 and 
the lowest $1A_{2}(n-\pi^{*})$ state is minimal, measuring at a 
mere 0.007 eV whereas the highest EE difference observed is for $2A_{1}
(n-\sigma^{*})$ state, measuring at a somewhat larger value of 0.192 eV. 
However, similar to the previous case, the actual magnitude of the states does not 
exhibit a uniform pattern with respect to the number of CSO involved. 
It is important to mention here that all the roots calculated here are 
dominated by single excitations except $1B_{2}$. In this case, a large
component of $S$ can be found with ($S_{p}:8b_{2}10a_{1}\rightarrow
9b_{1}9b_{1}$) resulting in a substantial influence from triples which
is somewhat missing from the description of CC3. A similar kind of 
trend can be observed in TZVP basis set as well. The lowest and highest
difference in EE between EOM-CCSD and CC3 is observed for $1A_{2}$
(0.019 eV) and $2A_{1}$ (0.239 eV) respectively. However, for EOM-iCCSDn
methodology, the lowest and highest error in EE with respect to CC3
comes around 0.005 eV (for $1A_{2}$) and 0.198 eV (for $2A_{1}$) respectively. 
Note that, both $2A_{1}$ and $3A_{1}$ states exhibit $\pi-\pi^{*}$ character, 
leading to an absence of a clearly defined $\pi-\pi^{*}$ valance transition.
It is also important to note here that, both $2A_{1}$ and $1B_{1}$ 
($\sigma-\pi^{*}$) states are prone to Rydberg-valance mixing.\cite{rydberg_valance}
Thus the accurate characterization of these states can be challenging due 
to the limitations of using cc-pVDZ basis sets, which are inadequate for 
capturing the high-lying transitions. Consequently, the estimated energy of 
these states may fall short of the true value, as evidenced by the best estimation 
of 9.1 eV for $1B_{1}$ state, according to reference.\cite{thiel} However, the 
utilization of TZVP basis set has shown some improvement in the accurate 
determination of these states compared to the cc-pVDZ basis set.

\begin{table*}[htbp]
\centering
\setlength{\tabcolsep}{2.5pt}
    \renewcommand{\arraystretch}{2.05}
    \begin{tabular}{ccc>{\columncolor[gray]{0.8}}c>{\columncolor[gray]{0.8}}c
    >{\columncolor[gray]{0.8}}ccc>{\columncolor[gray]{0.8}}c>{\columncolor[gray]{0.8}}c>
    {\columncolor[gray]{0.8}}c}
    \hline
    \hline
    \multirow{3}{*}{States} & \multicolumn{10}{c}{Basis Sets} \\\cline{2-11}
                            & \multicolumn{5}{c}{\textbf{cc-pVDZ}} 
            & \multicolumn{5}{c}{\textbf{TZVP}} \\
            \cmidrule(l){2-6} \cmidrule(l){7-11} 
                            & EOM-CCSD & CC3 & \multicolumn{3}{c}{EOM-iCCSDn}
            & EOM-CCSD & CC3 & \multicolumn{3}{c}{EOM-iCCSDn} \\
                            &     &    & (1,1) & (1,2) & (2,2)
            &     &          & (1,1) & (1,2) & (2,2) \\
    2$^{1}A_1$  & 9.947 & 9.711 & 9.875 & 9.871 & 9.903 & 9.769 & 9.530 & 9.701 & 9.698 & 
    9.728 \\
    3$^{1}A_1$  & 11.296 & 11.190 & 11.257 & 11.252 & 11.279 & 10.541 & 10.448 & 10.501 & 
    10.496 & 10.526 \\\hline
    1$^{1}B_1$ & 9.353 & 9.291 & 9.323 & 9.324 & 9.360 & 9.257 & 9.183 & 9.229 & 9.229 & 
    9.266 \\\hline
    1$^{1}B_2$ & 8.635 & 8.550 & 8.584 & 8.565 & 8.597 & 8.446 & 8.344 & 8.400 & 8.383 & 
    8.415 \\\hline
    1$^{1}A_2$ & 4.012 & 3.994 & 3.948 & 3.949 & 3.987 & 3.966 & 3.947 & 3.903 & 3.905 & 
    3.942 \\
    \hline
    \hline
    \end{tabular}
    \caption{Vertical excitation energies of Formaldehyde in different basis sets
    i.e. cc-pVDZ and TZVP. All the values given in the table are in eV. The values 
    in the parenthesis denotes the number of occupied and virtual orbitals considered
    to be included in CSO.}
    \label{tabmm3}
\end{table*}

\subsection{Cis-diazene ($N_{2}H_{2}$)}
The \textit{cis} isomer of the diazene ($N_2H_2$) molecule has been 
extensively studied in the field of excitation energy calculations. 
The optimized geometry used in these calculations was obtained from 
the Computational Chemistry Comparison and Benchmark Database (CCCBDB).\cite{cccbdb} 
Similar to the formaldehyde molecule, the cis-diazene molecule also 
belongs to the $C_{2v}$ point group and possesses four irreducible 
representations: $A_1$, $B_1$, $B_2$, and $A_2$.
In our study, we employed two different basis sets, namely cc-pVDZ 
and TZVP, for the calculation of excitation energies. For
the calculations, all the electrons were correlated. All the results for
different basis sets are enlisted in table \ref{tabmm4}. Excitation energies
predicted by EOM-CCSD and EOM-iCCSDn methodologies are compared against
the robust CC3 method. For the EOM-CCSD method using the cc-pVDZ basis set, 
the excitation energies oscillate in the range from 0.023 eV (for $1B_1$)
to 0.574 eV (for the $2A_1$ state) compared to the CC3 method. The 
excitation energies predicted by our EOM-iCCSDn method reside in between 
those of EOM-CCSD and CC3 for all states. Notably, for the $1B_1$ state, 
which is primarily governed by a $n-\pi^{*}$ excitation, our EOM-iCCSDn
methodology provided highly accurate excitation energies with a 
negligible difference of only 0.003 eV compared to CC3, whereas the 
largest error in excitation energy was observed for the $2A_1$ state 
($n-\sigma^{*}$) with a difference of 0.285 eV. It is important to mention 
here that all the states considered in our study are singlet in nature.
In contrast to previous cases, the magnitudes of excitation energies displayed 
a monotonic pattern with respect to the number of CSO involved for both 
the cc-pVDZ and TZVP basis sets. EOM-iCCSDn exhibits slightly better 
excitation energies when employing the CSO-(3,3) compared to CSO-(2,2). 
A similar trend in the EE is observed for the TZVP basis set as well, 
where the lowest and highest differences in EE predicted by EOM-CCSD with 
respect to CC3 were observed for the $1B_1$ state (0.023 eV) and 
the $1A_2$ state (0.206 eV), respectively. However, our EOM-iCCSDn method 
showed even more accurate excitation energies, with the lowest and 
highest errors compared to CC3 found for the $1B_1$ state 
($EE_{EOM-iCCSDn}- EE_{CC3}=0.001$ eV) and the $2A_1$ state (0.121 eV), respectively.
Consequently, our EOM-iCCSDn method demonstrates superior accuracy in 
predicting excitation energies compared to EOM-CCSD with respect to
robust CC3 method. This study highlights the potential of our methodology 
for accurate determination of excitation energies in theoretical 
investigations involving the cis isomer of the diazene molecule.
\begin{table*}[htbp]
\centering
\setlength{\tabcolsep}{6.5pt}
    \renewcommand{\arraystretch}{1.55}
    \begin{tabular}{ccc>{\columncolor[gray]{0.8}}c
    >{\columncolor[gray]{0.8}}ccc>{\columncolor[gray]{0.8}}c>
    {\columncolor[gray]{0.8}}c}
    \hline
    \hline
    \multirow{3}{*}{States} & \multicolumn{8}{c}{Basis Sets} \\\cline{2-9}
                            & \multicolumn{4}{c}{\textbf{cc-pVDZ}} 
            & \multicolumn{4}{c}{\textbf{TZVP}} \\
            \cmidrule(l){2-5} \cmidrule(l){6-9} 
                            & EOM-CCSD & CC3 & \multicolumn{2}{c}{EOM-iCCSDn}
            & EOM-CCSD & CC3 & \multicolumn{2}{c}{EOM-iCCSDn} \\
                            &     &    & (2,2) & (3,3)
            &     &          & (2,2) & (3,3) \\
    2$^{1}A_1$  & 9.414 & 8.840 & 9.257 & 9.125 & 8.726 & 8.556 & 8.703 & 8.677 \\\hline
    1$^{1}B_1$ & 3.840 & 3.817 & 3.847 & 3.820 & 3.744 & 3.721 & 3.750 & 3.722 \\\hline
    1$^{1}B_2$ & 6.912 & 6.870 & 6.894 & 6.874 & 6.505 & 6.436 & 6.489 & 6.472 \\\hline
    1$^{1}A_2$ & 6.748 & 6.564 & 6.548 & 6.549 & 6.643 & 6.437 & 6.435 & 6.434 \\
    \hline
    \hline
    \end{tabular}
    \caption{Vertical excitation energies of cis-Diazene in different basis sets
    i.e. cc-pVDZ and TZVP. All the values given in the table are in eV. }
    \label{tabmm4}
\end{table*}

\subsection{Trans-diazene ($N_{2}H_{2}$)}
Similar to the \textit{cis-}isomer, the \textit{trans-}isomer 
of diazene molecule has also been studied in the realm of 
excitation energies. The geometry of the molecule used here
was taken from CCCBDB database.\cite{cccbdb} Owing to the $C_{2h}$ point
group, the irreducible representations are $A_g,A_u,B_u$ and 
$B_g$. We have used cc-pVDZ and TZVP basis sets for the calculation
of excitation energies and all the results corresponding to
$A_g$,$B_u$ and $B_g$ symmetries are enlisted in table \ref{tabmm5}.
As previously, all the electrons were correlated for the calculations.
\begin{table*}[htbp]
\centering
\setlength{\tabcolsep}{6.5pt}
    \renewcommand{\arraystretch}{1.65}
    \begin{tabular}{ccc>{\columncolor[gray]{0.8}}c
    >{\columncolor[gray]{0.8}}ccc>{\columncolor[gray]{0.8}}c>
    {\columncolor[gray]{0.8}}c}
    \hline
    \hline
    \multirow{3}{*}{States} & \multicolumn{8}{c}{Basis Sets} \\\cline{2-9}
                            & \multicolumn{4}{c}{\textbf{cc-pVDZ}} 
            & \multicolumn{4}{c}{\textbf{TZVP}} \\
            \cmidrule(l){2-5} \cmidrule(l){6-9} 
                            & EOM-CCSD & CC3 & \multicolumn{2}{c}{EOM-iCCSDn}
            & EOM-CCSD & CC3 & \multicolumn{2}{c}{EOM-iCCSDn} \\
                            &     &    & (2,2) & (3,3)
            &     &          & (2,2) & (3,3) \\
    2$^{1}A_g$ & 7.731 & 7.670 & 7.709 & 7.683 & 7.195 & 7.114 & 7.182 & 7.159 \\\hline
    1$^{1}B_u$ & 7.750 & 7.692 & 7.729 & 7.707 & 7.401 & 7.312 & 7.384 & 7.367 \\\hline
    1$^{1}B_g$ & 3.289 & 3.277 & 3.304 & 3.276 & 3.263 & 3.249 & 3.276 & 3.247 \\
    \hline
    \hline
    \end{tabular}
    \caption{Vertical excitation energies of trans-Diazene in different basis sets
    i.e. cc-pVDZ and TZVP. All the values given in the table are in eV.}
    \label{tabmm5}
\end{table*}
We have compared and contrasted our EOM-iCCSDn methodology against
EOM-CCSD and CC3. In case of cc-pVDZ basis, the difference of EE
between EOM-CCSD and CC3 methods varies between 0.012 eV (for $1B_g$)
and 0.061 eV (for $1A_g$). The excitation energies predicted by EOM-iCCSDn
method follow the same overall trend and the values reside in between
EOM-CCSD and CC3. Significantly, for $1B_g$ state (primarily governed 
by $\pi-\pi^{*}$ excitation), EOM-iCCSDn method predicts excitation
energy quite accurately with an imperceptible difference of only 0.001
eV with respect to CC3. The largest error in EE compared to CC3 is
0.039 eV for $1A_g$ (governed by $n-\sigma^{*}$). It is important to
note here that all the states are dictated by single excitations.
From the values given in table \ref{tabmm5}, it is imperative to
demonstrate that the magnitudes of excitation energies exhibit a 
monotonic dependence on the CSO. In case of TZVP basis 
set, the highest and lowest error in EE for EOM-CCSD with respect 
to CC3 are 0.089 eV (for $1B_u$; $n-\sigma^{*}$) and 0.014 eV 
(for $1B_g$) respectively. However, our EOM-iCCSDn method is capable 
of predicting EE in a more accurate manner with the highest and 
lowest error in EE compared to CC3 comes around 0.072 eV (for $1B_u$)
and $EE_{EOM-iCCSDn}- EE_{CC3}= 0.002$ eV (for $1B_g$) respectively.

\subsection{Ethylene ($C_{2}H_{4}$)}
Another molecule that we have studied is Ethylene or Ethene.
The geometry used here for the calculation is taken from Thiel
test set.\cite{thiel} Ethene molecule belongs to the point group 
of $D_{2h}$ and owing to that, there are 8 irreducible representations 
possible i.e. $A_g$, $B_{3u}$, $B_{2u}$, $B_{1g}$, $B_{1u}$, $B_{2g}$,
$B_{3g}$ and $A_u$. In table \ref{tabmm6}, we have enlisted the excitation
energies corresponding to some of the symmetry states. We have employed
cc-pVDZ basis for the calculation and the values predicted by our
EOM-iCCSDn method is compared against EOM-CCSD and CC3. For all the
states discussed here, the difference in EE between EOM-CCSD and CC3
vary between 0.048 eV (for 1$B_{2g}$) to 0.474 eV (for 3$A_g$). For all
the states, the EE predicted by the EOM-iCCSDn methodology reside in between
EOM-CCSD and CC3 values; with the lowest and highest EE differences with
respect to CC3 are 0.015 eV (for 1$B_{2g}$) and 0.136 eV (for 2$B_{3g}$)
respectively. However, the magnitudes of excitation energy follow
monotonic dependence on the number of CSO. Note that, all the states
considered here are dominated by single excitations. However, for a 
few states, contrary to the EOM-CCSD prediction, a large magnitude of
$S$ is observed. For example, in case of 2$A_g$ and 3$A_g$ states, 
one finds a large component of $S$ ($S_{p}: 8b_{3u}10a_{g}\rightarrow 
12b_{2u}9b_{1g}$) along with a larger magnitude of $T_2$ ($8b_{3u}8b_{3u}
\rightarrow 9b_{1g}9b_{1g}$) which in turn incorporates a substantial effect 
from higher-order $T_3$ excitations which otherwise is missing 
from the CC3 calculations. Apart from that, 2$B_3g$ state also has a large
contribution from $S$ operator ($S_p: 8b_{3u}11b_{1u}\rightarrow
10a_{g}9b_{1g}$) along with a large component from $T_2$ ($7b_{3g}
8b_{3u}\rightarrow12b_{2u}9b_{1g}$). It is also important to note here
that this 2$B_{3g}$ state has a very high-lying excitation and may
be prone to rydberg-valance mixing. Thus from table \ref{tabmm6},
it can be inferred that our EOM-iCCSDn methodology can predict the
excitation energies in an accurate manner.
\begin{table}[htbp]
\centering
 \hglue 0.0cm
\setlength{\tabcolsep}{8pt}
\renewcommand{\arraystretch}{1.4}
\begin{tabular}{ccc>{\columncolor[gray]{0.8}}c>{\columncolor[gray]{0.8}}c}
\hline
\hline
Roots & EOM-CCSD & CC3 & \multicolumn{2}{c}{EOM-iCCSDn} \\
\hline
         &     &           & (2,2) & (4,4)\\
\cline{4-5} 
 2$A_g$ & 13.606 & 13.480 & 13.590 & 13.568 \\
 3$A_g$ & 15.196 & 14.008 & 14.130 & 14.126 \\
 4$A_g$ & 15.572 & 15.098 & 15.184 & 15.158 \\
\hline
 1$B_{1u}$ & 12.777 & 12.702 & 12.766 & 12.745 \\
 2$B_{1u}$ & 14.051 & 13.946 & 14.048 & 14.023 \\ 
\hline
 1$B_{2g}$ & 8.902 & 8.854 & 8.892 & 8.869 \\
\hline
 1$B_{3g}$ & 11.548 & 11.487 & 11.533 & 11.514 \\
 2$B_{3g}$ & 15.824 & 15.634 & 15.770 & 15.736 \\
\hline
 1$A_u$ & 11.804 & 11.587 & 11.664 & 11.658 \\
\hline
\hline
\end{tabular}
\caption{Vertical excitation energy for ethylene in the cc-pVDZ basis set. All 
the values given in the table are in eV.}
\label{tabmm6}
\end{table}

\newpage
\subsection{Nitrogen ($N_2$)}
Nitrogen ($N_2$) molecule is another system that has been studied significantly
in the area of excited state calculations. While the point group of 
$N_2$ is $D_{\infty h}$, for the purpose of computational implementation, 
we will consider the highest abelian group, $D_{2h}$ corresponding to the 
$D_{\infty h}$ point group. In our application, we have considered only
two symmetry states i.e. $\Pi_{u}$ (corresponds to $B_{3u}$ and
governed by $\sigma^{*}-\pi^{*}$ excitation) and $\Pi_{g}$ (corresponds 
to $B_{2g}$ and governed by $\pi-\pi^{*}$) of $N_2$ which are dominated by 
single excitations. We have calculated the excitation energies of these 
two states in several basis sets i.e. cc-pVDZ, cc-pVTZ, TZVP, cc-pCVDZ 
and cc-pCVTZ. For all the calculations, all the electrons were correlated.
All the results for all the basis sets are enlisted in table \ref{tabmm7}.
\begin{table*}[htbp]
\centering
\setlength{\tabcolsep}{17pt}
\renewcommand{\arraystretch}{1.7}
\begin{tabular}{cccc}
\hline
\hline
Basis Sets & Methods & \multicolumn{2}{c}{States} \\
\hline
           &         &  $\Pi_u$ & $\Pi_g$ \\
\cline{3-4} 
 \multirow{4}{*}{\textbf{cc-pVDZ}} & CC3 & 13.820 & 9.680 \\
                                   & EOM-CCSD & 14.039 & 9.727 \\
                                   & \cellcolor{lightgray} EOM-iCCSDn (2,2) 
                                   & \cellcolor{lightgray} 13.851 
                                   & \cellcolor{lightgray} 9.704 \\
                                   & \cellcolor{lightgray} EOM-iCCSDn (3,3) 
                                   & \cellcolor{lightgray} 13.754 
                                   & \cellcolor{lightgray} 9.701 \\
\hline
 \multirow{4}{*}{\textbf{cc-pVTZ}} & CC3 & 13.573 & 9.559 \\
                                   & EOM-CCSD & 13.828 & 9.620 \\
                                   & \cellcolor{lightgray} EOM-iCCSDn (2,2) 
                                   & \cellcolor{lightgray} 13.637 
                                   & \cellcolor{lightgray} 9.592 \\
                                   & \cellcolor{lightgray} EOM-iCCSDn (3,3) 
                                   & \cellcolor{lightgray} 13.548 
                                   & \cellcolor{lightgray} 9.589 \\
\hline
 \multirow{4}{*}{\textbf{TZVP}}    & CC3 & 13.644 & 9.629 \\
                                   & EOM-CCSD & 13.865 & 9.692 \\
                                   & \cellcolor{lightgray} EOM-iCCSDn (2,2) 
                                   & \cellcolor{lightgray} 13.683 
                                   & \cellcolor{lightgray} 9.667 \\
                                   & \cellcolor{lightgray} EOM-iCCSDn (3,3) 
                                   & \cellcolor{lightgray} 13.586 
                                   & \cellcolor{lightgray} 9.664 \\
\hline
\multirow{4}{*}{\textbf{cc-pCVDZ}} & CC3 & 13.834 & 9.689 \\
                                   & EOM-CCSD & 14.061 & 9.740 \\
                                   & \cellcolor{lightgray} EOM-iCCSDn (2,2) 
                                   & \cellcolor{lightgray} 13.873 
                                   & \cellcolor{lightgray} 9.716 \\
                                   & \cellcolor{lightgray} EOM-iCCSDn (3,3) 
                                   & \cellcolor{lightgray} 13.777 
                                   & \cellcolor{lightgray} 9.714 \\
\hline
\multirow{4}{*}{\textbf{cc-pCVTZ}} & CC3 & 13.577 & 9.554 \\
                                   & EOM-CCSD & 13.844 & 9.620 \\
                                   & \cellcolor{lightgray} EOM-iCCSDn (2,2) 
                                   & \cellcolor{lightgray} 13.653 
                                   & \cellcolor{lightgray} 9.592 \\
                                   & \cellcolor{lightgray} EOM-iCCSDn (3,3) 
                                   & \cellcolor{lightgray} 13.564 
                                   & \cellcolor{lightgray} 9.589 \\
\hline
\hline
\end{tabular}
\caption{Vertical excitation energies of Nitrogen in different basis sets. 
    All the values given in the table are in eV. The values in the parenthesis 
    denote the number of occupied and virtual orbitals considered to be included 
    in CSO.}
\label{tabmm7}
\end{table*}
We have compared and contrasted the values of excitation energy
predicted by EOM-iCCSDn methodology against EOM-CCSD and robust
CC3 methods. We have also employed two different Sets of CSO. 
In all the basis sets, for $\Pi_g$ state, EOM-iCCSDn values reside
in between EOM-CCSD and CC3 values. It can be shown that the values
can systematically be improved upon including a larger number of 
orbitals in CSO.

\section{Summary and future direction}
\label{conclude}
We have therefore developed, implemented, and benchmarked our
iCCSDn methodology that is capable of inducing effects of high-rank 
excitations in the EOM framework. We have developed a novel projective algorithm 
followed by a sufficiency condition towards the
development of the working equations and demonstrated that 
the computational overhead is at par with the parent iCCSDn method. 
We have employed our methodology on a series of moderately to strongly correlated
molecules in and compared the results against robust methods.
Whereas for a number of molecules, we have shown that there is a non-monotonic 
dependence of excitation energy on the number of contractible orbitals,  
we have also shown a few examples where the excitation energy can 
systematically be improved upon the inclusion of a higher number of contractible
orbitals. Through the series of pilot numerical applications,
we have shown that our EOM-iCCSDn methodology can faithfully
reproduce the correct qualitative as well as quantitative
behavior compared to other allied methodologies, but at a
much less computational scaling. Since we have tools for 
calculating ground state energetics of large chemical systems as
well as excited state energetics, it would be straightforward
to assimilate both methodologies in order to determine the local excitation energy 
of large embedded or extended systems.



\section*{Acknowledgements}
The authors thank Prof. Debashis Mukherjee, India for all 
the stimulating discussions during the development of 
the theory. AC thanks Industrial Research and Consultancy
Center (IRCC), IIT Bombay, for research fellowship. AC thanks
Dibyendu Mondal for stimulating discussions and valuable insights.

\section*{Author Declarations}
\subsection*{Conflict of Interest:}
The authors have no conflict of interest to disclose.

\subsection*{DATA AVAILABILITY}
The data that support the findings of this study are
available from the corresponding author upon 
reasonable request.

\bibliography{literature}

\end{document}